\documentclass{PoS}

\PoS{PoS(LAT2005)204}

\title{Heavy quark propagators with improved precision using domain decomposition\thanks{Preprint: HU-EP/05/41, SFB/CPP-05-42, SHEP-0528}}

\ShortTitle{Heavy quark propagators with improved precision}

\author{\speaker{Andreas J\"uttner}\\
        University of Southampton, School of Physics and Astronomy,\\
	Highfield, Southampton, SO17 1BJ, United Kingdom\\
        E-mail: \email{juettner@phys.soton.ac.uk}}

\author{Michele Della Morte\\
	Humboldt Universit\"at zu Berlin, Institut f\"ur Physik,\\
	Newtonstrasse~15, D-12489 Berlin, Germany\\
	E-mail: \email{dellamor@physik.hu-berlin.de}}

\abstract{
We show that, in four dimensions, the quark propagator is affected by round-off
errors for large values of the quark mass $am$ and the time extent $T/a$ even 
when double precision arithmetics is used to compute it.
We introduce a definition of the solver residual which is sensitive to the 
problem and apply a Schwarz alternating procedure to compute the propagator in 
a number of sub-domains in the time direction.
The effectiveness of the method is demonstrated in a numerical computation of the free one-dimensional Dirac propagator.
}

\FullConference{XXIIIrd International Symposium on Lattice Field Theory\\
		 25-30 July 2005\\
		 Trinity College, Dublin, Ireland}
\usepackage{amssymb,amsmath}
\usepackage{epsfig}
\usepackage{psfrag}
\newcommand{\be}{\begin{equation}}
\newcommand{\ee}{\end{equation}}
\newcommand{\bea}{\begin{equation}\begin{array}}
\newcommand{\eea}{\end{array}\end{equation}}
\newcommand{\bdm}{\begin{displaymath}}
\newcommand{\edm}{\end{displaymath}}
\newcommand{\rba}{\begin{array}}
\newcommand{\rea}{\end{array}}
\newcommand{\bi}{\begin{itemize}}
\newcommand{\ei}{\end{itemize}}
\newcommand{\bver}{\begin{verbatim}}
\newcommand{\everbatim}{\end{verbatim}}

\begin{document}

\section{Introduction}
Lattice simulations with a  relativistic $b$-quark 
 will become feasible with the next generation of super computers
 \cite{Boyle:2005gf,Bodin:2001hn,Ukawa:2005hn}. 
At least in the quenched approximation, these machines should allow for lattice 
sizes $L/a={\rm O}(100)$ so that the conditions  $am_b<1$ and $L\simeq 1-2\,$fm
can be fulfilled at the same time 
(see \cite{Juttner:2005tb,Rolf:2003mn} for studies approaching this regime). 
Among other things, such computations will provide  clear tests of effective field
theories like HQET~\cite{Eichten:1989zv} or of alternative formulations for 
relativistic heavy quarks 
on the lattice  \cite{El-Khadra:1996mp,Aoki:2001ra}. 
Depending on the outcome, further evidence can be produced
 supporting the use of these theories, especially in view of computations
with light dynamical flavors, where lattice sizes suitable for the inclusion of
relativistic (quenched) heavy quarks are far to come.

On the numerical side simulations in that regime may be affected by round-off
errors.
The lattice quark propagator is usually computed numerically  by employing CG-type 
algorithms to solve the system of linear equations $D\psi = \eta$
for $\psi$, where the matrix $D$ is some discrete representation of the Euclidean 
Dirac operator.
Defining the \emph{time slice norm} of a Dirac vector $\psi(t,\vec{x})$ on the time
slice $t$ as 
\be
|\psi|_t=\sqrt{\sum\limits_{\vec{x},a,\alpha}\left(\psi^{a}_\alpha(\vec{x},t)\right)^\ast \psi^a_\alpha(\vec{x},t)}\, ,
\ee
(greek indices represent spin, roman indices colour)
we argue that in situations where 
\be\label{whennecessary}
{\min\limits_t |\psi|_t\over \max\limits_t|\psi|_t} \simeq {\rm arithmetic
 \;precision},
\ee
the solver residual 
\be\label{conventionalres}
r=\sqrt{\sum\limits_t |\eta-D\psi|_t^2/\sum\limits_t|\eta|^2_t}
\ee 
becomes an un-reliable indicator for convergence and the solver itself fails to 
produce sensible results. Both effects are due to accumulated round-off errors.

We propose an algorithm with the potential to overcome these problems. 
We suggest decomposing the time direction of the lattice into a sufficient 
number of adjacent or overlapping domains to avoiding the situation in 
eq.~(\ref{whennecessary}) within each domain. By applying the Schwarz alternating 
procedure \cite{Schwarz:1890,Saad} to these domains, we are able to recursively 
construct the solution over the whole lattice in a controlled way.

We first present a numerical example for the case of free Wilson fermions in the 
4-dimensional QCD Schr\"odinger Functional \cite{Sint:1994un} where a CG-type 
solver produces unreliable results although proper convergence is indicated by 
the residual in eq. (\ref{conventionalres}). 
We then review the proposed algorithm and demonstrate its efficiency considering as 
an example the 1-dimensional Dirac equation. We compare our numerical results to an 
\emph{exact} reference solution for the propagator in this model, computed with Mathematica. The term 
\emph{exact} refers to the fact that within this framework one can vary the
arithmetic precision even beyond double precision and thereby gain confidence in
 the numerical solution.
As reported in \cite{Luscher:2003qa,Luscher:2003vf,Luscher:2004rx}, the Schwarz 
alternating procedure can of course also be applied to the fully interacting
 4-dimensional theory.
\section{The problem in 4 dimensions}
\begin{figure}
\begin{center}
\begin{minipage}{.45\linewidth}
\psfrag{xlabel}[c][bc][1][0]{$T/a$}
\psfrag{ylabel}[c][c][1][0]{$\log_{10}r$}
\begin{center}
\epsfig{scale=.45,file=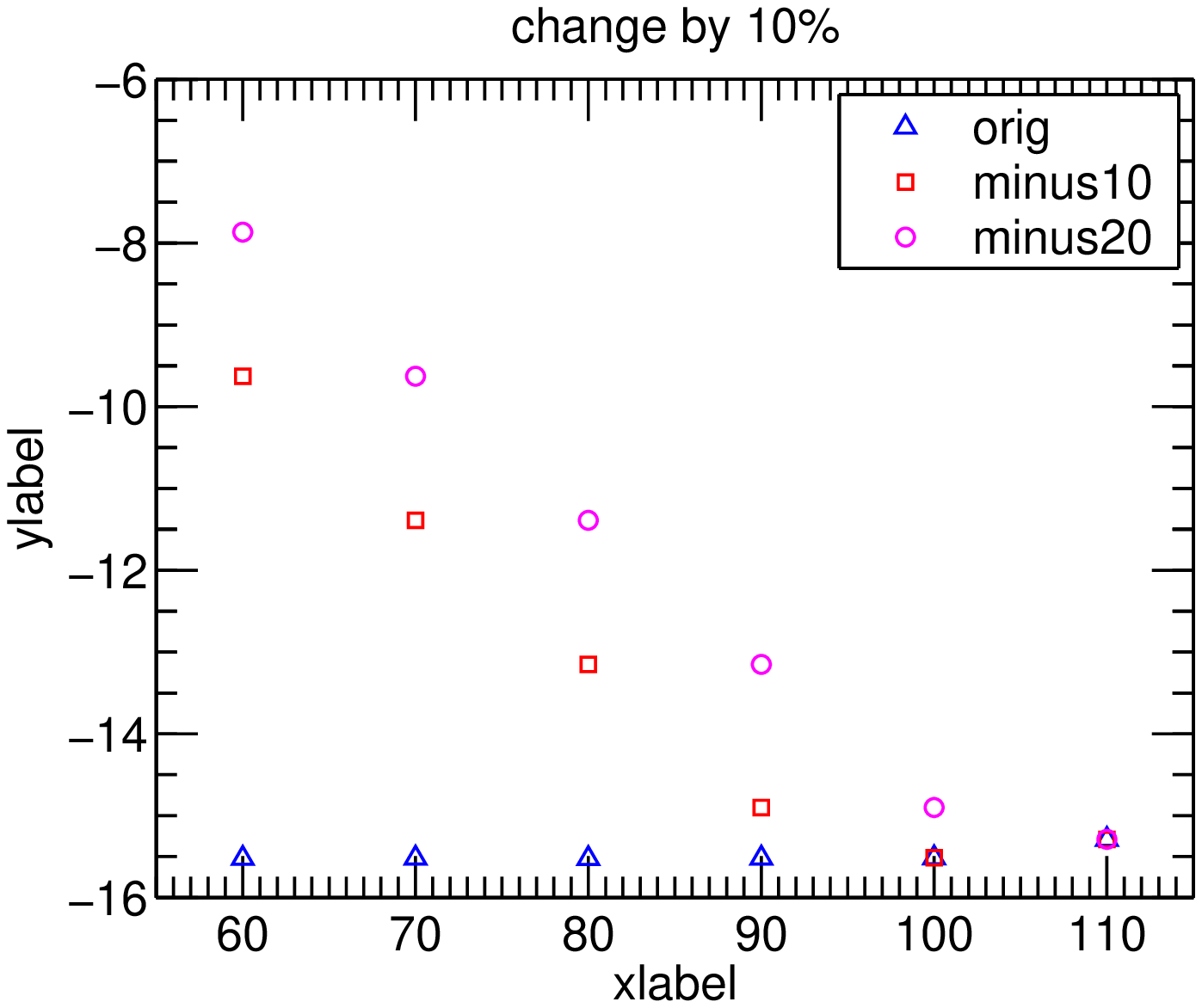}
\end{center}
\end{minipage}
\begin{minipage}{.45\linewidth}
\psfrag{xlabel}[c][bc][1][0]{$T/a$}
\psfrag{ylabel}[c][c][1][0]{$\log_{10}r$}
\begin{center}
\epsfig{scale=.45,file=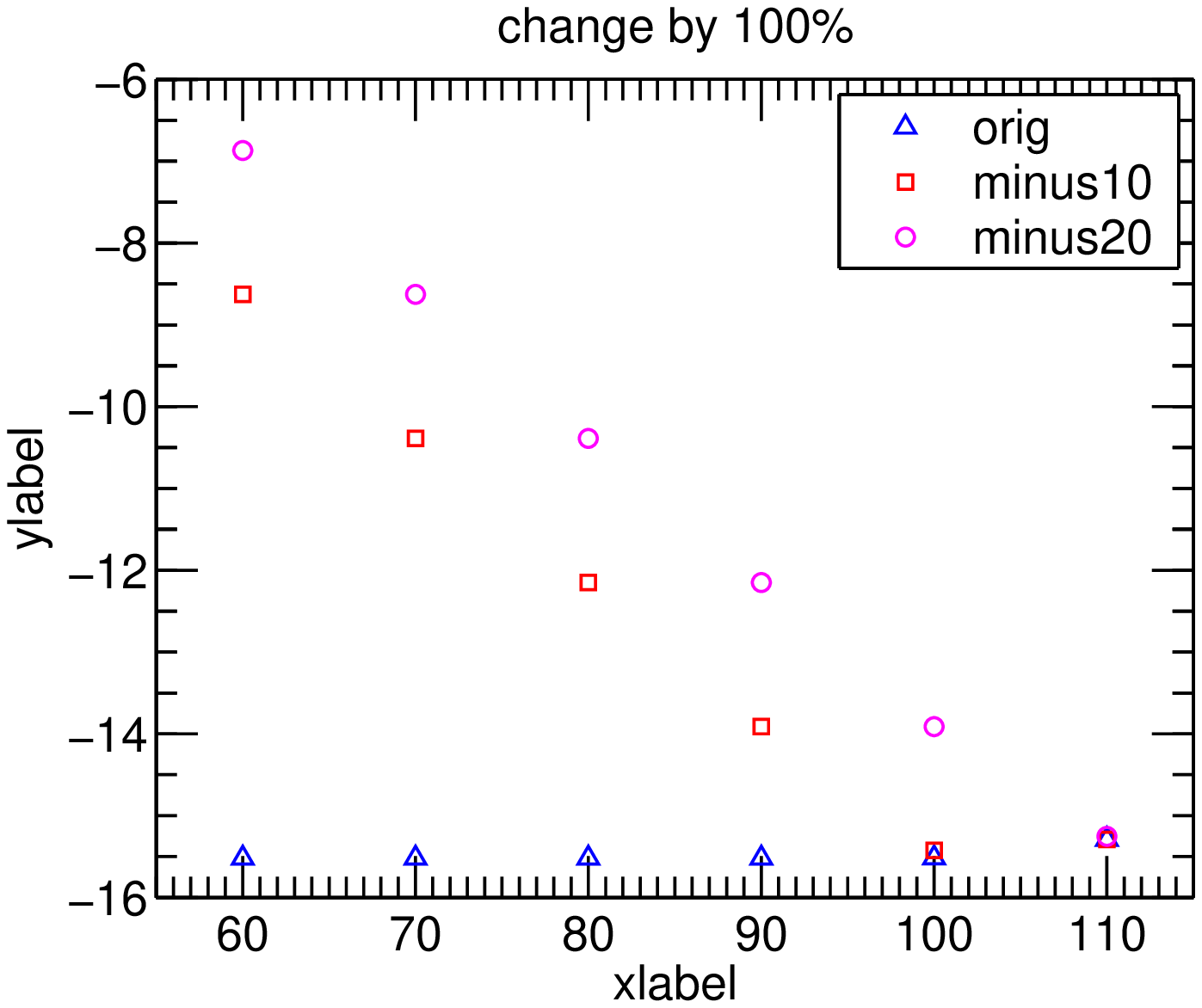}
\end{center}
\end{minipage}
\caption{BiCGstab solver residual $r$ for the solution of the free Wilson Dirac
 propagator on a $12^3\times T$-lattice with $T=\{60,70,80,90,100,110\}$ after
 convergence (triangles) and after having changed the solution on the last 10
 time slices (squares) and on the last 20 time slices (circles) by $+10\%$ and
 $+100\%$ respectively.}\label{4Dresults}
\end{center}
\vspace{-.4cm}
\end{figure}
We illustrate the problem with a numerical study in the QCD Schr\"odinger Functional using the double precision version of the
 MILC code \cite{MILC}. The parameters were  $am\simeq 0.5$ ($\kappa=0.{111111}$),
 $L/a=12$, and $T/a=\{60,70,80,90,100,110\}$ ($a=1$ in the following) and we used
 a unit gauge background. Using the stabilised bi-conjugate gradient algorithm
 (BiCGstab) \cite{Frommer:1994vn}, we solved $D\psi=\eta$ for $\psi$,
 which corresponds to a column of the quark propagator. As is common practise, we
 used $r\le 10^{-15}$ for the stopping criterion.

As a test of the solver residual and of proper convergence for large $t$, we
 changed the solution $\psi(t,\vec{x})$ by +10\% (or +100\%), once for $t\ge T-10$
 and once for $t\ge T-20$ for all the values of $T$ mentioned above and then
 recomputed $r$. Figure \ref{4Dresults} shows the results. The triangles represent
 the achieved solver residual for each choice of $T$ and the squares and circles
 represent the residual $r$ after having changed the solution for  $t\ge T-10$ and
 $t\ge T-20$, respectively. We see that above a certain time-extent 
 $T$ of O($100$) the residual
 ceases to be sensitive to a change of the solution by 10\% (or 100\%). In this
 situation the computed solution $\psi$ cannot be considered correct.

\section{The 1-dimensional Dirac operator}
The problem we observed in 4 dimensions is also present in 1-di\-men\-sio\-n.
To illustrate this we implemented the equation $D\psi=\eta$ in MATLAB
 with $D$ given by the 1-dimensional free Wilson lattice Dirac operator with periodic boundary conditions,
\be\label{nonhermdirac}
D^{(1)}_{xy}\equiv \delta_{xy} - \kappa\left[\delta_{y,x+1}(1+\gamma_1)+\delta_{y,x-1}(1-\gamma_1)\right],
\ee
where $\kappa=1/(2m+2)$ is the hopping parameter. We solved for $\psi$ using a
 stabilised BiCG algorithm in double precision. Since we implemented
 the Dirac equation on the torus, we expect the problem to appear at around twice 
the time extent observed in the Schr\"odinger functional. 
Indeed, for $T=180,\;m=0.5$ and $r=10^{-15}$ the solution
vector varies in time by more orders of magnitude than can be represented by the
 arithmetic precision and vanishes exactly for the 10 central lattice points. 
One therefore expects the solution to be wrong for a time interval larger than 10
 time slices around $T/2$. 

\vspace{-0.2cm}
\section{The (multiplicative) Schwarz alternating procedure (SAP)}
We now give our proposal to circumvent the problem. We decompose the time-direction of the lattice into a number of domains, such that the solution is expected to decay by fewer orders of magnitude than covered by the arithmetic precision within each domain.

\begin{figure}
\begin{center}
\psfrag{7}[c][l]{$\Lambda$}
\psfrag{8}[c][l]{$\partial\Lambda^\ast$}
\psfrag{9}[c][l]{$\partial\Lambda$}
\psfrag{10}[c][l]{$\Lambda^\ast$}

\epsfig{scale=1.5,file=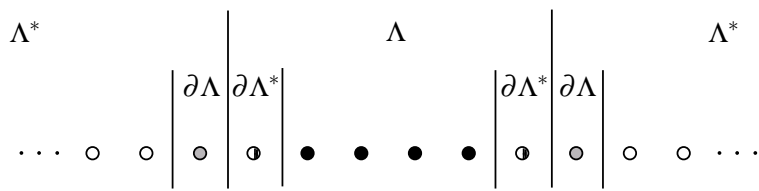}
\end{center}\caption{Domain of size $L_1=6$. The black dots represent the bulk of the domain $\Lambda$, the interior boundary $\partial \Lambda^\ast$ is represented by the half-filled dots and the exterior boundary $\partial \Lambda$ is represented by the grey circles. The complement of $\Lambda$ is $\Lambda^\ast$.}\label{domainpic}
\vspace{-0.3cm}
\end{figure}

We adopt the notation given by L\"uscher \cite{Luscher:2003qa} and briefly review some basic definitions. 
We decompose the problem into $n_{\rm dom}$ non-overlapping domains.  
Each domain $\Lambda$ has an interior boundary $\partial \Lambda^\ast$ and an exterior boundary $\partial \Lambda$ (cf. figure \ref{domainpic}).
The position-space Dirac operator may be written in the form
\be
D = \left(
\rba{ll}
D_\Lambda			&D_{\partial \Lambda}\\
D_{\partial \Lambda^\ast}	&D_{\Lambda^\ast}\\
\rea
\right),
\ee
where the matrices $D_\Lambda$ and $D_{\Lambda^\ast}$ act on the domain $\Lambda$ and its complement $\Lambda^\ast$, respectively. The off-diagonal matrices $D_{\partial \Lambda}$ and $D_{\partial \Lambda^\ast}$ contain those interactions that couple $\Lambda$ to the adjacent domains. 

Following \cite{Luscher:2003qa}, the algorithm we propose visits
 each of the $n_{\rm dom}$ domains in successive sweeps and updates the current
 approximation $\psi$ to the solution of $D\psi=\eta$ 
according to 
\be
\psi^\prime = \psi + D_\Lambda^{-1}(\eta - D\psi).
\ee 
Here we take $\psi=0$ as the initial guess.
We introduce the domain based stopping criterion\footnote{Here $|\cdot|_\Lambda$ is the Dirac norm restricted to the domain $\Lambda$.}\\
\be
r_{\rm dom} =  \max\limits_{\Lambda}\left\{{\sqrt{ |D\psi-\eta|^2_\Lambda/  |\psi|_\Lambda^2}}\right\}\le 10^{-15}.
\ee
Note that we normalise with respect to the solution vector since one usually uses a $
\delta$-function as source $\eta$.

\vspace{-0.3cm}
\section{Numerical results}

\begin{figure}
\begin{center}
\begin{minipage}{.45\linewidth}
\psfrag{xlabel}[c][bc][1][0]{$t/a$}
\psfrag{ylabel}[c][c][1][0]{\footnotesize$\rm log_{10}$(relative error)}
\epsfig{scale=.4,file=bildla/sap_m0.5_L120}
\end{minipage}
\begin{minipage}{.45\linewidth}
\vspace{-2mm}
\psfrag{xlabel}[c][c][1][0]{ $t/a$}
\psfrag{ylabel}[c][t][1][0]{\footnotesize${\rm log}_{10}( |D_\Lambda \delta -(\eta-D \psi)|_t/|\delta|_t)$} 
\epsfig{scale=.4,file=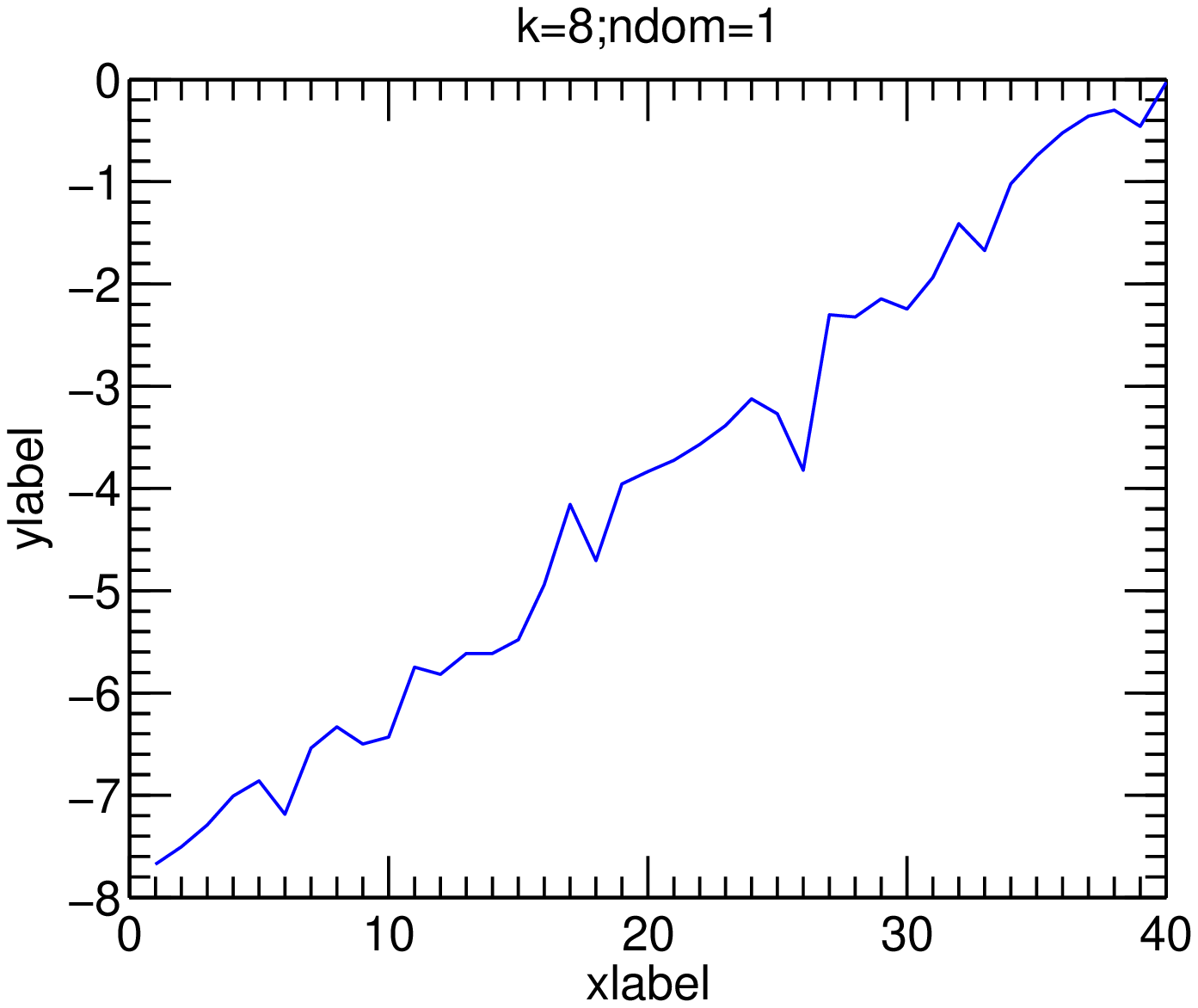}
\end{minipage}\\
\end{center}
\caption{Left: Example of solver accuracy for $n_{\rm dom}=3, m=0.5$ and $T=120$ 
(vertical lines indicate the domain-borders). The blue lines (red circles) 
represent the relative  error of the result obtained by the BiCG (SAP) solver
with respect to the \emph{exact} solution. 
Right:  Time-slice residual for the problem $D_\Lambda^{-1}(\eta-D\psi)=\delta$ 
after the 8th sweep of the SAP solver over the 1st domain ($T=120$).
}\label{figresults1}
\end{figure}

\vspace{-0.2cm}
We first computed $\psi$ on the whole lattice in 1 dimension by means of Fourier
 transformation in Mathematica. Since this software allows for arbitrary numerical
 precision we could thereby obtain an \emph{exact} reference solution. We then 
implemented the SAP solver in MATLAB, where all further numerical tests have been 
performed.
Within each sub-domain the current update for the solution $\psi$ is computed
 using a BiCG solver which runs until convergence. As an example we discuss the
 case $T=120$ and $m=0.5$.  The SAP solver takes about 10 times more iterations
 than the conventional (unreliable) BiCG with a global stopping criterion 
like in eq.~(\ref{conventionalres}).
Notice that  the  matrix $\times$ vector operations needed in the
 SAP  solver clearly involve smaller matrices.
 However, for heavy quarks the condition
 number of the Dirac matrix is not very large and the main issue is rather the
 precision. The results for both algorithms are illustrated in 
figure \ref{figresults1} in terms of the time-slice relative
error with respect to the \emph{exact} solution.
Two comments are in order :
\bi
\item Without domain decomposition the solution $\psi$ deviates strongly from the \emph{exact} solution despite alleged proper convergence of the solver indicated by the residual $r$.
\item The circles in fig. \ref{figresults1} show the relative error for the solution computed with the method suggested in this work. Notably it stays at the desired level over the whole time extent of the lattice, indicating uniform convergence. 
\ei

\vspace{-0.26cm}
\section{Conclusions}
We have given numerical evidence that conventional CG-solvers run into round-off problems when the lattice volume and the quark mass are large. In particular the solver residual $r$ given in eq. (\ref{conventionalres}) is misleading in these cases.
We have performed preliminary tests of an algorithm and a residual based on the Schwarz alternating procedure that do not suffer from these problems.

In contrast to the conventional solver, the algorithm we suggest converges to a constant precision over the whole lattice. Still one might expect the local residual to grow within each domain. This is indeed visible when the problem $D_\Lambda^{-1}(\eta-D\psi)=\delta$ is considered (see figure \ref{figresults1}, right plot).

The parameter range where the algorithm applies is complementary to that 
for conventional CG-solvers for small quark masses on the one hand and to that for the procedure
based on the hopping parameter expansion for very large quark masses 
\cite{Thacker:1990bm} on the other hand. The algorithm suggested here should in fact
allow a more precise assessment of the range of quark masses where the latter method is
applicable.

The implementation in 4 dimensions should be straightforward. In QCD a  gain 
in performance could presumably be achieved by starting with the
 free propagator as initial guess. It would be very interesting to investigate
 the influence of overlapping domains on the solver performance.

{\bf Acknowledgements:} We warmly thank Jonathan Flynn and Ulli Wolff for their
 helpful comments and for reading the manuscript. This work was supported by the
 SFB/TR 09 of the Deutsche Forschungsgesellschaft and the PPARC grant
 PPA/G/O/2002/00468.
\bibliography{proceeding}
\bibliographystyle{JHEP}
\end{document}